# The roads not taken: empty waves, wavefunction collapse and protective measurement in quantum theory

**Peter Holland**
Green Templeton College
University of Oxford
Oxford OX2 6HG
England

Email: peter.holland@gtc.ox.ac.uk

> *Two roads diverged in a wood, and I–*
> *I took the one less traveled by,*
> *And that has made all the difference.*
> 　　　　　　　　　　　　Robert Frost (1916)

**1 The explanatory role of empty waves in quantum theory**

In this contribution we shall be concerned with two classes of interpretations of quantum mechanics: the epistemological (the historically dominant view) and the ontological. The first views the wavefunction as just a repository of (statistical) information on a physical system. The other treats the wavefunction primarily as an element of physical reality, whilst generally retaining the statistical interpretation as a secondary property. There is as yet only theoretical justification for the programme of modelling quantum matter in terms of an objective spacetime process; that *some* way of imagining how the quantum world works between measurements is surely better than none. Indeed, a benefit of such an approach can be that 'measurements' lose their talismanic aspect and become just typical processes described by the theory.

　　　In the quest to model quantum systems one notes that, whilst the formalism makes reference to 'particle' properties such as mass, the linearly evolving wavefunction $\psi(x)$ does not generally exhibit any feature that could be put into correspondence with a localized particle structure. To turn quantum mechanics into a theory of matter and motion, with real atoms and molecules comprising particles structured by potentials and forces, it is necessary to bring in independent physical elements not represented in the basic formalism. The notion of an 'empty wave' is peculiar to those representatives of this class of extended theories which postulate that the additional physical element is a corpuscle-like entity or point particle. For clarity, we shall develop the discussion in terms of a definite model of this kind whose properties are well understood and which it is established reproduces the empirical content of quantum mechanics: the de Broglie-Bohm theory, a prominent representative of the class of ontological interpretations (Holland, 1993). Here, material physical systems are postulated to comprise two components: a physically real wave (described by $\psi(x)$) governed by Schrödinger's equation, and a point particle that is guided along a track $x(t)$ by the wave (according to the law $m\dot{x} = \nabla S$ where $S$ is the quantal phase) but does not participate in the latter's dynamics (one can extend the model to include a back-reaction of the particle on the wave in a way that is compatible with quantal predictions (Holland, 2006) but this is not needed here). The position of the particle is the 'observable' of the theory. Note that this dualistic theory of matter discerns, and attributes ontological significance to, features of the wavefunction – such as energy and force – that may not be meaningful in other ontological interpretations (which therefore may be incommensurable).

If the $\psi$-wave is incident on a beam-splitter and evolves into two spatially disjoint components, the particle will enter only one of them and the other, by virtue of not containing the particle, will be 'empty'. It is only in this sense that we shall say a wave is empty - it still propagates energy and momentum of the field throughout space and has the potential to subsequently act on its associated particle if it is finite in a domain where the latter passes (the energy-momentum is only indirectly observable through the effects of the wave on the particle).

To illustrate the active role of an empty wave we recall how it contributes to the de Broglie-Bohm description of the two-slit experiment (Holland, 1993, Sec. 5.1). Referring to Fig.1, a wave $\psi(x)$ incident on a beam-splitter $B$ splits into two packets $\psi_1$ and $\psi_2$ which separate sufficiently so that they do not appreciably overlap before being recombined in the vicinity of a screen $P$. Two distinct routes, 1 and 2, are then available to a particle $x$ passing through the interferometer. If the particle is detected at a point $X$ above the symmetry axis $A$ we know that it traversed route 1 because the single-valuedness of the wavefunction forbids the crossing of paths. Hence, after the splitting and prior to the recombination, $\psi_2$ is an empty wave. But the point $X$ may lie in a region not accessible to a particle guided by $\psi_1$ alone, that is, in the case where $\psi_2$ is absent (this may be arranged, for example, if $\psi_1$ possesses nodes where $\psi_2$ is finite). Hence, the empty wave $\psi_2$ has had a physical effect in bringing about an observable change in the state of the corpuscle: when $\psi_2$ is present (absent) the particle can (cannot) land at $X$.

Notice that the claim that the empty wave has acted physically is a retrospective inference - we argue that it must have so functioned prior to the detection of the corpuscle in order that the result obtained could actually occur. Naturally, the empty wave concept has meaning only within the model of quantum motion we have employed to explain the functioning of the interferometer; the interference phenomenon itself does not prove the 'reality' of the empty wave.

It would clearly be advantageous if the historical dispute between the epistemological and ontological viewpoints could be made an empirical issue. In this piece we shall examine the impact of the empty wave concept on this problem. We first emphasize the theoretical merits of the empty wave in enabling avoidance of the wavefunction collapse hypothesis (Sec. 2) and in supplying conceptual precision in the application of quantum mechanics, with particular reference to an example where protective measurements have been used in path detection (Sec. 3). We then go on to address the problem of how the reality of an empty wave might be demonstrated by its effect on other systems, and advance general arguments against this possibility (Sec. 4). However, these arguments are not conclusive and we describe how an alternative perspective in probing the empirical implications of empty waves is provided by the notion of protective measurement (Sec. 5).

**2 Measurement: empty waves *vs*. wavefunction collapse**

The empty wave concept extends easily to a many-particle system where it is a key characteristic of the configuration space description. An analogue of a beam splitter in configuration space will create a spectrum of waves and the system point will distinguish one of them if they are non-overlapping packets. Note that the physical particles composing the system point need not be located nearby in three-dimensional space.

A drawback of the epistemological interpretation is that it entails the hypothesis that the wavefunction 'collapses' at some stage in a measurement process as the knowledge of the 'observer' regarding the state of a system changes, a notion that is hard to formulate unambiguously and consistently (e.g., in relation to relativity). One of the virtues of the de Broglie-Bohm approach is that it provides a coherent account of measurement that, in particular, dispenses with the problematic collapse hypothesis through the use of empty waves.

The measurement problem of quantum mechanics arises when one attempts to attribute definite outcomes to processes devoted to discovering information on a quantum system (Holland, 1993, Chap. 8). The measurement of an observable represented by an operator $\hat{A}$ associated with a system having a coordinate $x$ is customarily modelled by an impulsive interaction generated by the Hamiltonian $H = f\hat{A}(-i\hbar\partial/\partial z)$ where $z$ is the coordinate of the apparatus and $f$ is a constant. At first the system and apparatus are non-interacting so the total initial state is $\Psi_0(x,z) = \psi_0(x)\phi_0(z)$ where $\psi_0(x) = \sum_a c_a \psi_a(x)$ is a superposition of eigenstates of $\hat{A}$, and $\phi_0(z)$ is the initial apparatus state (assumed to be a localized packet). The impulsive interaction acts as a beam splitter in configuration space generating a spectrum of macroscopically distinct apparatus states each correlated with an individual eigenfunction. If the period of interaction is $T$ we obtain

$$\Psi(x,z,T) = \sum_a c_a \psi_a(x,T) \phi_a(z,T) \tag{2.1}$$

where $\phi_a(z,T) = \phi_0(z - faT)$ represents a set of non-overlapping outgoing apparatus packets. These packets are in turn coupled to many-body packets so that their separation is amplified to the macroscopic scale. Each packet corresponds to a possible outcome of the measurement. But the state is a superposition of outcomes and, in order to extract a definite result from the superposition, the hypothesis is invoked in the epistemological interpretation that the state (2.1) 'collapses' into one of the summands, say the $a$th, with probability $|c_a|^2$:

$$\sum_{a'} c_{a'} \psi_{a'}(x,T) \phi_{a'}(z,T) \to \psi_a(x,T) \phi_a(z,T) \tag{2.2}$$

(after normalization). This transformation is not described by the unitary evolutionary law of quantum mechanics (Schrödinger's equation) and suggestions for how it might come about have ranged from the intervention of an observer who becomes aware of the outcome to modifications of the Schrödinger equation. But, even if it is assumed that it does actually take place, the notion of collapse does not in itself solve the measurement problem. For, to infer the outcome of the measurement, the pointer of the apparatus must be assigned a location whose variation during the interaction can be determined unambiguously. In contrast, according to its usual interpretation, the wavefunction attributed to the apparatus determines just the statistical frequency of measurement results. The wavefunction does not itself offer a description of an autonomous moving object. One may attempt to address this difficulty by invoking the feature that $\phi_0(z)$ is sharply peaked about a spacetime orbit, that is, by making some kind of literal identification of the packet with the particle. Then one is tacitly shifting the interpretation of the wavefunction towards an ontological view, but not in a clearly consistent way - the eventual diffusion of the packet, or the possibility of splitting it into disjoint parts, mean the 'particle' does not remain localized, for instance.

Another option is that the projection (2.2) does not take place. Rather, the correct wavefunction remains (2.1), so that all terms in the superposition continue to be finite, but one is selected as representing the outcome of the measurement because it carries some special attribute. This is the thesis of the de Broglie-Bohm model.

In an ensemble of particle systems the probability density of presence in the initial state is $|\Psi_0(x,z)|^2$. Then, in the measurement, one of the outgoing summands is singled out because the de Broglie-Bohm system point ($x(t),z(t)$) enters it (i.e., it occupies the region where the summand is finite). All the other packets are then empty. In particular, the outcome of the measurement is the position $z(t)$. Since the outgoing packets are non-overlapping, from the standpoint of the *particles* the transformation (2.2) does in effect occur, even though the other $\psi_a$s and $\phi_a$s are still finite (but empty). The Born probability formula follows since over an ensemble the particle $x$ enters the $a$th packet with relative frequency $|c_a|^2$. Within this approach, the entire measurement process may be treated by applying the usual linear, unitary Schrödinger equation, and the single concept of particle trajectory enables one to both avoid the collapse postulate and to solve the problem of the definiteness of the pointer (and object) position. We shall return to the issue of distinguishing the epistemological and ontological views in this context in Sec. 5.

**3 The art in quantum mechanics: path detection and conceptual precision**

*3.1 Theory of path detection*

The de Broglie-Bohm theory is particularly suited to analyzing the interplay between the observation of interference effects and the determination of the spacetime path of a quantum system. Path determination in such situations often requires establishing that the system lies within a particular spatial region, rather than locating it using a precision position measurement. For example, in an interferometric context the path may lie within one of two distinguishable beams traversing a device. In that and other settings path detection may be achieved by entangling the system of interest with another system that has the characteristics of a detector, i.e., that has macroscopically distinguishable states each of which is uniquely coupled with one of the available beams. Here we discuss and attempt to resolve a controversy that has arisen surrounding the application of the ideas of particle trajectory and empty wave in a 'which-path' context. It is shown that, if these ideas are applied correctly according to the principles of the de Broglie-Bohm theory, they provide a coherent and uncontroversial account of the functioning of these devices.

An example of such a device arises in the spin $\frac{1}{2}$ version of the EPR experiment. Referring to Fig. 2, a source *S* generates a pair of oppositely moving particles (with magnetic moments) in a singlet state. The particle with spatial coordinate *y* (the detector) passes through a Stern-Gerlach magnet on the right-hand side oriented in any direction. If *y* is detected in the upper of the two beams emerging from the magnet (path 1, spin up) then we may infer without further investigation that particle *x* on the left-hand side pursues the lower beam (path 2), if it subsequently passes through a Stern-Gerlach device oriented in the same direction. Notice that in this example the determination of the path of *x* has been achieved via a remote local action of the right-hand magnet. Indeed, in this example, the particles never come near each other during the detection process (and the two may be located as far apart as one desires so long as the entangled state is preserved). This is possible because the local action on the magnetic moment of *y* does not exhaust the dynamical influence of the right-hand magnetic field on the particles, which is mediated also by the wavefunction. The latter carries information on the local interaction, which is thereby transmitted to *y* (causing it to move along path 1 or 2) and (nonlocally) to *x*. It is essential to appreciate that, in this sort of example, the motions of the particles *x* and *y* are correlated not because they act upon one another directly as would be expected for two classically interacting particles but because they are each guided by the wavefunction that carries an imprint of the entire experimental context.

This remote action, whereby a detector locates the de Broglie-Bohm trajectory of a particle with which it does not directly (classically) interact, or even come near, was understood in the context of nonlocal EPR correlations in the 1980s (see Holland, 1993, Sec. 11.3 and references therein). However, a conceptually similar example of path detection in interferometry published subsequently has occasioned some (unfounded) disquiet, as we now discuss.

The following is an elaboration of the discussion in Holland (1993, Sec. 8.8). Suppose in Fig. 1 we introduce a device in path 1 having wavefunction $\phi(y)$ and coordinate *y(t)* (see Fig. 3). The purpose of this device is to couple $\psi_1(x)$ and $\phi(y)$ so that the distinct states of the latter allow us to learn along which beam, 1 or 2, *x* traverses the interferometer. Initially, the total wavefunction is

$$\Psi(x,z) = [\psi_1(x) + \psi_2(x)]\phi(y) \rightarrow \gamma(x,y) + \psi_2(x)\phi(y) \text{ during interaction in path 1}$$

$$\rightarrow \psi_1(x)\phi'(y) + \psi_2(x)\phi(y) \qquad (3.1)$$

after the interaction in path 1, which is assumed to leave $\psi_1(x)$ essentially unaltered. It is required that the initial and final detector states are disjoint in the space of their argument (*y*), $\phi \cap \phi' = 0$, in order that an unambiguous reading is obtained. Then the two configuration space summands to which they contribute are disjoint. Thus, if *y* is found to lie in the excited state $\phi'(y)$, *x* must lie in $\psi_1(x)$. The possible outcomes are as follows:

$$\left.\begin{array}{l} y_0 \rightarrow y \in \phi'(y) \Rightarrow x \in \psi_1(x) \\ y_0 \rightarrow y \in \phi(y) \Rightarrow x \in \psi_2(x). \end{array}\right\} \qquad (3.2)$$

The outcome in each trial is fixed uniquely by the initial positions $x_0$, $y_0$ and the total wavefunction $\Psi$. What can we conclude about the path the particle $x$ took through the interferometer?

To bring out how the inference drawn from the meter reading $y$ depends on the total wavefunction, we consider two possible final wavefunctions, corresponding to two different experiments that differ in the location of the detecting screen: case (a) when $\psi_1$ and $\psi_2$ overlap, and case (b) after $\psi_1$ and $\psi_2$ have recombined and, following their natural evolution, passed through one another and separated. Then, from (3.2), when the detector is excited, $x$ may be deduced to pass along path 1 in case (a) and along path 2 in case (b).

It will be noted that, in case (b), $x$ does not pass along the path (1) where the detector is located, that is, an empty wave is associated with the excitation, and the detector locates the particle in a region remote from it. We also find in case (b) that when the detector is unexcited the particle passes through it. Although the details of the devices in Figs. 2 and 3b differ, they display the same feature of remote detection and for the same reason: correlated motion of the two particles induced by the total wavefunction that expresses the entanglement of the detector and object and develops into a sum of two disjoint product states in configuration space.

*3.2 Realism vs. surrealism*

Englert, Scully and co-workers (Englert *et al*., 1992) have sought to use these features of scenario (b) to argue that the de Broglie-Bohm theory is not a 'realistic' description because the trajectories 'may be macroscopically at variance with the observed track of the particle'. They present their argument using a Stern-Gerlach interferometer having a detector in each arm but the simpler set-up used in Fig. 3b with scalar wavefunctions and a single detector (suggested by Dewdney *et al*. (1993)) suffices. Their key claim is that for a path detection to occur a detector must fire due to a local interaction between it and the particle whose path is desired, which must be at the detector's location. In an alternative example (Aharonov *et al*., 1999), employing a protective measurement (see Sec. 5) to effect a path detection[1], they suggest that, in the de Broglie-Bohm theory, the momentum transfer to the meter that is involved in the measurement process should be accounted for by the action of the particle $x$ and that, since the particle $x$ in the case they consider does not pass near the relevant point where the interaction 'takes place', it cannot have this physical effect. Insofar as their criteria for path detection are not obeyed by the de Broglie-Bohm theory, these authors introduce an artistic metaphor and assert that the trajectories are 'surreal'. What they seem to mean by this mode of expression is that the trajectories are 'wrong'. They suggest that the de Broglie-Bohm model needs to specify additional criteria to determine when a legitimate path detection is effected.

Of course, were it the case that a correct description of quantum path detection entailed detectors functioning in the way Englert, Scully and co-workers claim, that is, as involving purely local interactions that reveal the particle trajectory at the location of the detector, this would be an awkward circumstance for the de Broglie-Bohm description. What justifies the claim that a quantum trajectory theory must display such characteristics?

---

[1] Whether a protective measurement can be assimilated to a position measurement has been questioned by Drezet (2006).

In analyzing this question the merits of the de Broglie-Bohm approach, in prompting an examination of how language is employed in quantum theory, come to the fore. In the application to protective measurement, Englert, Scully and co-workers justify their claims through appeal to terms like 'common quantum sense' and 'well localized interaction' as if they are unproblematic components of a clear conceptual framework within which it is legitimate to judge the meaning of the de Broglie-Bohm theory. But the true situation is the inverse of this: the de Broglie-Bohm theory provides the means to assess the worth of a traditional discourse that comprises an extraordinarily vague amalgam of words and concepts tacked on to the quantum formalism. That is, the terms commonly used in quantum mechanics are in fact highly problematic, in particular because they are not part of, or mapped onto, a clear ontology. The purpose of the de Broglie-Bohm theory is precisely to address these shortcomings by providing a consistent framework within which the meanings of terms commonly used in quantal discourse may be assessed. For example, the 'conventional view' expounded by these authors that the excitation of a detector functioning through a 'well localized interaction' is in itself sufficient to claim that a particle passed through it is unfounded unless supplemented by a physical model consistent with quantum mechanics that allows us to formulate criteria in terms of which it is meaningful to draw such an inference. What is the model of a 'particle' for which it can be meaningfully asserted that it 'passes'? The conventional view fails to satisfy physicists' natural desire for an unambiguous ontology and for want of an alternative its adherents often slip into a classical discourse for which there is no justification in this context and which, moreover, cannot be consistent. In his paper entitled 'Do Bohm trajectories always provide a trustworthy physical picture of particle motion?', Scully (1998) answers in the Abstract 'No. When particle detectors are included particles do not follow the Bohm trajectories as we would expect from a classical type model.' And there is the nub of the issue: these critics want interpretations of quantum theory to conform to classical conceptions. A pre-quantum notion of interaction comprising a local exchange of momentum (that has not been proven to be consistent with quantum mechanics) is being invoked to judge a theory (that is proven to be consistent) that indicates how a quite different nonclassical notion of 'interaction' is necessary if the particle trajectory is to be deployed in a quantum context. In fact, what these authors claim is the 'observed track' according to their classical model may indeed not be the actual track based on a quantum model.

As emphasized in relation to the example of Fig. 2, in the de Broglie-Bohm model the 'interaction' is defined not just by the form of the Hamiltonian but is an action mediated by the configuration space wavefunction, which implies (nonlocally) correlated motions in three-dimensional space. Thus, *local Hamiltonians have nonlocal effects*. The statement of Aharonov *et al*. (1999) that '…an interaction between the particle and the meter occurs undoubtedly…' is the kind of loose language objected to above. When one tries to make this notion precise, as in the de Broglie-Bohm theory, it is seen that it is not a meaningful statement. Rather, one must say that there is an action by the wavefunction on the two corpuscles, which causes them to evolve in a correlated manner so that from the path of one we may infer the path of the other.

Even if particle *x* travels through the detector when the latter fires, as happens with case (a), there is still no direct interaction between *x* and *y*, and *the excitation has not occurred because of the passage*. Indeed, giving significance to the excited as opposed to the unexcited state of the detector is misleading, for in both states one can make an inference as to the path traversed. The fact that the assertion 'click =

detection of passing particle' is generally unfounded does not so much signal a flaw in the de Broglie-Bohm theory as alert us to the subtlety of the quantum theory of detection that it reveals, in particular that the question of whether or not the particle traverses the detector is *irrelevant* to the issue of path detection. The interaction embodied in (3.1), governed by Schrödinger's equation, occurs whether the wave is empty or not. According to this theory, a path detector *never* directly records the coordinates of the particle 'measured' or its 'passage'.

Indeed, the arbitrary requirement that in a scenario deemed to be one of 'path detection' the detected particle must pass in the vicinity of the detector, regardless of the prevailing quantum state, would lead over an ensemble to distributions of readings at variance with those implied by $|\Psi|^2$. In contrast, the de Broglie-Bohm description honours the quantal predictions faithfully. And no additional criteria are needed to specify when a path detection occurs, beyond the reading of the meter *y*.

In the light of the above we can also assess the analysis of Dewdney *et al.* (1993) who have described the detector as being 'fooled' in case (b). In fact, this is somewhat misleading since in all cases the detector performs its function of indicating the particle route; it is no more fooled in this case than in case (a) or in the example of Fig. 2.

Distant actions of local interactions are at the heart of the explanatory framework of the de Broglie-Bohm theory and examples abound already in the single particle case. For example, a particle approaching an infinite barrier (the local interaction Hamiltonian) will be reflected without touching it. This happens because the wavefunction carries information about the local potential (the barrier) to distant points and guides a corpuscle located there. This is not 'surreal'; it just shows how quantum theory transcends classical mechanism.

There *is* art in the de Broglie-Bohm picture but it is a subtle, non-classical realism based on a concept of particle interaction for which there is no obvious precedent in pre-quantum physics. The latter aspect is relevant to quantum path detection because of the use of entanglement as a resource. Englert, Scully and co-workers have neglected this feature and hence their criticism is unfounded. One may not care for this aspect of the trajectory theory but to cite it as a blemish in the de Broglie-Bohm description the critic must propose a consistent alternative ontology. As indicated above, in this connection it is not legitimate to invoke as a benchmark 'standard quantum mechanics' whose lack of precision was a key motivation for the development of the casual theory in the first place. It has, in fact, often been the lot of the de Broglie-Bohm interpretation to be reproached for seeking to return to classical conceptions (by employing trajectories) only to be faulted for not being classical enough (the trajectories do not do what the critic wants).

**4 Evidence for empty waves: retrodiction *vs.* prediction**

*4.1 A general argument against the observability of empty waves*

How could the reality of an empty wave be demonstrated? We shall explore here the view that what is desired is a means of manipulating such an entity and its interactions so as to measurably alter the future course of systems that may be potentially influenced by it in a *predictable* way. Two potential methods present themselves. The validity of either method would contradict the hypothesis of wavefunction collapse. A first method is to bring the empty wave back to influence its own associated corpuscle once we have established the latter's path. Applied to the measurement procedure described in Sec. 2, this would entail reversing the process to achieve overlap of the outgoing macroscopic apparatus packets, a formidable technical challenge. A second method is to try to manipulate the empty wave so as to influence *another* independent wave-particle composite, thus increasing the size of the relevant Hilbert (and configuration) space. An argument has been given (Holland, 1993, Sec. 8.8) that for a general class of interactions the latter method does not allow one to infer the reality of an empty wave, at least according to the criterion of predictability stated above. We now recall this demonstration.

Suppose an initial packet $\psi(x)$ containing a particle with coordinate $x(t)$ is split into two packets $\psi_1(x)$ and $\psi_2(x)$ that subsequently separate so that eventually they do not appreciably overlap (Fig. 4). The particle will join one or other of the packets. Suppose that $\psi_1$ interacts with a detector having wavefunction $\varphi(z)$ and coordinate $z(t)$ that can measure the position $x$, and that subsequently $\psi_2$ interacts with some other system having wavefunction $\xi(w)$ and actual location $w(t)$. Initially, the total wavefunction is

$$\Psi(x,w,z) = [\psi_1(x) + \psi_2(x)]\xi(w)\varphi(z). \qquad (4.1)$$

After the first interaction has commenced the wavefunction is non-factorizable:

$$\Psi(x,w,z) = \alpha(x,z)\xi(w) + \psi_2(x)\xi(w)\varphi(z). \qquad (4.2)$$

The function $\alpha(x,z)$ is entangled in its configuration space and evolves into a superposition of sharply peaked functions of $x$ (with $z$-dependent coefficients, cf. (2.1)). If the corpuscle $x$ lies in $\psi_1$ it will be found in a region where one of these functions is finite and we then know $\psi_2$ is an empty wave (since $\psi_1 \cap \psi_2 = 0$). Now, if we bring in the second interaction, between $x$ and $w$, the wavefunction (4.2) becomes

$$\Psi(x,w,z) = \alpha(x,z)\xi(w) + \beta(x,w)\varphi(z). \qquad (4.3)$$

Since the two summands in (4.3) do not overlap, the system point is in one of them. But we have detected that $x(t)$ is in the first summand so we conclude that $w(t)$ cannot be in the second. Hence, because the spatial structure of the function relevant to the motion of the particle $w$ ($\xi(w)$) is unaltered, $\psi_2(x)$ has no observable effect on the particle behaviour and we cannot prove the hypothesis of empty waves.

As in the case of Fig.1, what is forbidden is the prediction of an effect of the empty wave once $x$ is detected; we can infer only the past action. As regards the future evolution, empty packets interact only with other *empty* packets - the particles of the other systems are in the same configuration space packet as the particle of interest. All the empty packets do indeed interact with one another and mutually modify their behaviour but this is unobservable since no particles are involved (recall that in this theory the outcomes of experiments are the positions of particles).

This feature of the configuration space dynamics is consistent with experience. If empty waves could really alter the measurable properties of systems, experiments would be constantly perturbed by background noise caused by extraneous $\psi$-fields and it would be hard to justify the assumption of relative autonomy in which it is legitimate to isolate segments of matter and ignore their environment.

That it is legitimate to draw retrospective inferences, which involve physical elements not included in the predictive apparatus of quantum mechanics, is a feature of some conventional presentations. Thus, Heisenberg (1930) admitted the possibility of reconstructing a trajectory in an interferometer. In both that case and in the examples of empty wave behaviour considered here, making correct retrodictions requires adopting a consistent theory of quantum particle motion, or results in contradiction with the statistical distributions of quantum mechanics may be obtained (Holland, 1993, Sec. 8.4.2).

*4.2 A stronger argument*

In the argument just presented against the observability of an empty wave, $\psi_1(x)$ ($\psi_2(x)$) is non-empty (empty) throughout the process. One may envisage more complex scenarios, such as are encountered with beams propagating through an interferometer that separate and then recombine, where these roles may be reversed for periods of the process, prior to the final stage when $\psi_2(x)$ is empty. This introduces the possibility that, in a period when $\psi_1(x)$, say, is empty, we can introduce an interaction in its domain with a detector $y$ so that an outcome of the entire process is that the latter's state alters observably. This raises an issue as to whether this intermediate event could imply evidence for the reality of an empty wave. We shall show that our result remains valid in this more general situation.

In fact, just this circumstance of reversed roles and intermediate interaction with a detector is implicit in the arrangement of Fig. 3b and we shall present our proof with reference to that. Including a detector $z$ that interacts with $x$ in the region where $\psi_1(x)$ is finite, the wavefunction (3.1), in the period after the waves $\psi_1(x)$ and $\psi_2(x)$ have passed through one another and no longer overlap, evolves into

$$\Psi(x,w,y,z) = \alpha(x,z)\phi'(y)\xi(w) + \psi_2(x)\xi(w)\phi(y)\varphi(z). \qquad (4.4)$$

If $x$ is detected by the detector $z$ then it lies in the first summand and, since the summands do not overlap, so does $w$ (and $y$). Therefore, at this final stage the packet $\psi_2(x)$ is an empty wave and an interaction between $x$ and $w$ (which induces $\psi_2(x)\xi(w) \to \beta(x,w)$) will not influence the future behaviour of $w$, as argued in the case of Fig. 4. Hence, we reaffirm our contention that a wave we know to be empty cannot observably influence the future behaviour of another physical system.

As anticipated above, the novel element in this example is that over the course of the process the status of $\psi_1(x)$ changes from empty, when it is party (along with the potentials involved in the interaction) to the transformation $\phi(y) \to \phi'(y)$, to non-empty, when it subsequently interacts with $z$ (in cases where the latter makes a detection). It has been argued by Hardy (1992) in connection with a similar arrangement[2] that, since the then-empty wave $\psi_1$ is a (partial, in the case he considers) cause of the observable change in *y*, this is evidence that empty waves can 'manifest their reality'. However, according to our analysis this will not be so. It is true that the empty wave $\psi_1$ contributed to *y*'s change in state, but because this action occurred *before* the *x*-detection by *z* we cannot impute any greater reality to the empty wave here than was possible in the case of, say, Fig.1. Suppose we include in the description of the process depicted in Fig. 1 the detection at the screen *P*, and couple the detector to a light bulb that glows if the detector registers a detection at *X*. Then we may predict that, when the particle is detected at *X*, the bulb will glow and this change in its state is caused by the past action of the empty wave $\psi_2$. The prediction embodied in the change in *y* in Fig. 3b is of a similar type. We only know $\psi_1$ was empty, in the period when the change occurred, after the *z*-detection. But by that stage there is no empty wave involved; the relevant empty wave then is $\psi_2$. It is therefore a retrospective inference that $\psi_1$ caused *y*'s variation. The empty wave certainly contributes to the theoretical account of how the results come about (as analyzed in the de Broglie-Bohm approach) but if the experiment was performed and the quantum predictions confirmed (hardly in doubt) this would not provide evidence for the reality of empty waves.

Although the above results are general and not restricted in the systems to which they apply, there is a caveat: they rest on the assumption that the final interaction between *x* and *w* must be localized in a domain of configuration space remote from the region where the first summand is finite. Although this is a natural assumption, it is overly restrictive if we aspire to a comprehensive assessment of all conceivable observable effects of empty waves. Another option is that the system *w* may interact with both summands in *xz*-space in a way that maintains their disjointness yet imparts to the first (system-point containing) summand an observable influence depending on the second summand. This possibility is examined next.

**5 Evidence for empty waves: protective measurement**

Progress in attempts to demonstrate an ontological aspect of the wavefunction came in 1993 when Aharonov and co-workers (Aharonov and Vaidman (1993), Aharonov *et al*. (1993) and Anandan (1993); for reviews and clarifications see Dass and Qureshi (1999) and Gao (2013)) showed how a suitably adapted adiabatic interaction described by quantum mechanics provides a scheme to measure the expectation values of operators pertaining to a system without appreciably disturbing its quantum state. These interactions are therefore called 'protective' measurements. In certain circumstances this provides a technique for 'measuring the wavefunction' of a single system as an extended object (this is not to be confused with the possibility of reconstructing the wavefunction from a statistical ensemble of conventional measurements (Holland, 1993, Sec. 8.7)). Aharonov *et al*. infer from this procedure, which reveals a property possessed by a single system prior to the measurement, evidence for the ontological character of the wavefunction. Here we shall point out how the protective measurement protocol, applied in the context of the de Broglie-Bohm model, potentially provides additional support for the ontological viewpoint, by devising a scheme that could distinguish between the empty wave and wavefunction collapse hypotheses. This application was first suggested by Holland (1994).

---

[2] The difference between the layouts of Fig. 3b and that of Hardy (1992) is the inclusion in the latter of a beam-splitter in the region where $\psi_1$ and $\psi_2$ overlap. This introduces an additional interaction so that $\psi_1$ is only a partial cause of the transformation $\phi \rightarrow \phi'$. Fig. 3b is free of this complication.

We first summarize the theory of protective measurements. Let the initial moment of time be $t = T$ and consider two interacting systems, an object and a measuring apparatus, with initial wavefunctions $\alpha(x,T)$ and $\beta(y,T)$, respectively. Denote by $\hat{B}$ the operator pertaining to the object whose expectation value is to be measured. Then in the protective interaction envisaged by Aharonov and co-workers, the interaction Hamiltonian is $H = g(t)y\hat{B}$ and the initial combined state $\Phi(x,y,T) = \alpha(x,T)\beta(y,T)$ evolves adiabatically at time $t$ into:

$$\Phi(x,y,t) = \alpha(x,t)\beta(y,t)\exp\left[-(i/\hbar)\int_T^t g(t)y\langle\hat{B}\rangle dt\right]. \tag{5.1}$$

Here $g(t)$ is a function characterizing the interaction with $\int_T^t g(t)dt = 1$, and $\alpha(x,t)$ and $\beta(y,t)$ are the wavefunctions obtained under free evolution of the two systems. It will be observed that this is still a product state in that the variables $x$ and $y$ have not become entangled. In particular, the object state is undisturbed by the interaction. In contrast, the state of the apparatus has acquired a phase factor, which implies an observable change in its momentum, depending on the expectation value $\langle\hat{B}\rangle = \langle\alpha(t)|\hat{B}|\alpha(t)\rangle$. Hence, information on the state $\alpha(x,t)$ can be gleaned from the apparatus by measuring the change in its momentum. For example, we can choose $\hat{B} = |x_0\rangle\langle x_0|$ so that $\langle\hat{B}\rangle = |\alpha(x_0,t)|^2$ and the shift in momentum is given by $\int_T^t g(t)|\alpha(x_0,t)|^2 dt$, the time-averaged square of the wave amplitude at the point $x_0$.

If $\alpha$ is known to be a non-degenerate energy eigenstate, but is otherwise unknown, it is possible to use this scheme to measure it for all values of its argument (up to a gauge transformation) by suitable choices of $\hat{B}$. It is to this case that the notion of 'measuring the wavefunction' of a single system using the protective technique really applies. The method may also be applied to general states but there are two caveats: (a) the full Hamiltonian that functions during the protective process depends on the state (Aharonov *et al.*, 1993), which implies that we must first know $\alpha$ before we can investigate it, and (b) that investigation reveals results about time averages of functions of the wavefunction rather than instantaneous values. So, in the general case the protective scheme provides a way to confirm empirically our time-averaged prior information. Our application of the protective technique below falls into this category; when the protective interaction commences the wavefunction is the result of a known state preparation procedure (a conventional measurement process). The aspect of the protective process that is important here is that any finite portion of the wavefunction of interest (obtained by varying its argument) has a discernible effect on a measuring device.

By extending the range of the label *x* the scheme may be generalized in a straightforward way to provide a protective measurement of a many-particle system, which again may be applied in principle to any wavefunction. The formula (5.1) remains the same if a single-component observable $\hat{B}$ is measured. It has been argued that this extension supports the attribution of ontological significance to the wavefunction in configuration space (Anandan, 1993).

We propose to apply this method to the wavefunction (2.1) that results from a typical conventional measurement in the case where it is assumed the collapse (2.2) does not occur, that is, when (2.1) comprises the set of empty waves generated by the measurement interaction in addition to the one corresponding to the actual outcome. The initial wavefunction is then the function (2.1); this is the wavefunction to which the protective interaction is applied (so that we replace *x* above by *x* and *z*). Let us suppose that the configuration point (*x,z*) of the de Broglie-Bohm model lies in the *a*th summand of $\Psi(x,z,T)$ and that we determine this fact, and hence the location of corpuscle *x*, by registering *z*. Then the other summands are finite but empty from the moment the summands separate and remain so independently of the registration of *x*. Hence, we may attempt to apply the technique of Aharonov *et al.* to measure functions of the finite, empty components of the total wavefunction and so provide empirical support for their reality. To this end, fix attention on the $a'$th component, $a' \neq a$, choose a point $(x_0, z_0) \in \psi_{a'}(x,t)\phi_{a'}(z,t) \neq 0$ (so that $\psi_{a''}(x_0,t)\phi_{a''}(z_0,t) = 0$ for all $a'' \neq a'$), and let the operator pertaining to the object (here the first object plus first detector) be $\hat{B} = |z_0\rangle|x_0\rangle\langle x_0|\langle z_0|$. Then, from (5.1) we obtain

$$\Psi(x,y,z,t) = \left[\sum_a c_a \psi_a(x,t)\phi_a(z,t)\right]\beta(y,t)$$
$$\times \exp\left[-(i/\hbar)y\int_T^t g(t)|\psi_{a'}(x_0,t)\phi_{a'}(z_0,t)|^2 dt\right] \quad (5.2)$$

To test whether in the first (conventional) measurement the wavefunction has really collapsed then requires observing the momentum of the detector *y*, which in the ontological interpretation has shifted by an amount depending on the finite amplitude of the empty wave $\psi_{a'}(x,t)\phi_{a'}(z,t)$.

This technique provides an alternative to the first method of testing for empty waves mentioned at the start of Sec. 4 in that it is not necessary to get an empty wave to overlap with the packet containing the system point. It is also not necessary to seek to observe the superposition of outgoing states since we need only select one for attention. There are, however, formidable difficulties of implementation. Some of the practicalities of the protective scheme, such as contamination by entanglement and the problem of measuring the variation in the meter's state, have been discussed in the above references, particularly by Dass and Qureshi (1999). A significant issue for the above proposal is that, for the selected outgoing configuration space packet, the protective scheme is being applied to a *macroscopic* object (through the detector coordinate *z* and its coupling to further many-body systems). A possible arena in which to apply the scheme is that of the cases studied in Sec. 4. During a protective measurement, the wavefunction (4.2) evolves into

$$\Psi(x,w,z) = \alpha(x,z)\xi(w)e^{iwp(\psi_2,\varphi)/\hbar} + \psi_2(x)\varphi(z)\xi(w)e^{iwp(\psi_2,\varphi)/\hbar} \qquad (5.3)$$

rather than (4.3). The system point remains in the first summand in (5.3) and the detector coordinate *w* acquires a momentum *p* depending on the empty wave amplitude $\psi_2(x)\varphi(z)$. A similar result is obtained with the wavefunction (4.4).

It should be pointed out that a less stringent interaction than that of the protective scheme may suffice for the purpose of observing the effect of an empty wave. The key attribute used here is the ability to probe the quantum state as an extended object; it may be permissible to allow the probed state to be modified by some 'quasi'-protective interaction, for example.

## 6 Conclusion

We have considered an aspect of the problem of how evidence may be gained to support the contention that a quantum system has a particle component. The empty wave, a concomitant of the particle model, is a useful theoretical element but in generic situations its influence can at best be inferred retrospectively. This is consistent with the fact that empty waves do not generally disturb physical systems. But there are exceptions and we have described how this issue could potentially be brought into the experimental arena when the special conditions of the procedure used in a protective measurement are satisfied. Distinguishing collapse from non-collapse models is theoretically feasible but technically demanding and the challenge is to find a practical implementation of the protective method.

## References

Aharonov, Y. and Vaidman, L. (1993). Measurement of the Schrödinger wave of a single particle. *Phys. Lett. A*, **178**, 38.
Aharonov, Y., Anandan, J. and Vaidman, L. (1993). Meaning of the wave function. *Phys. Rev. A*, **47**, 4616.
Aharonov, Y., Englert, B.-G. and Scully, M.O. (1999). Protective measurements and Bohm trajectories. *Phys. Lett. A*, **263**, 137; Erratum. *Phys. Lett., A*, **266**, 216 (2000).
Anandan, J. (1993). Protective measurement and quantum reality. *Found. Phys. Lett.*, **6**, 503.
Dass, N. D. H. and Qureshi, T. (1999). Critique of protective measurements. *Phys. Rev. A,* **59**, 2590.
Dewdney, C., Hardy, L. and Squires, E.J. (1993). How late measurements of quantum trajectories can fool a detector. *Phys. Lett. A*, **184**, 6.
Drezet, A. (2006). Comment on: "Protective measurements and Bohm trajectories" [Phys. Lett. A 263 (1999) 137]. *Phys. Lett. A*, **350**, 416.
Englert, B.-G., Scully, M.O., Süssmann, G. and Walther, H. (1992). Surrealistic Bohm trajectories. *Z. Naturforsch.*, **47a**, 1175.
Frost, R. (1916). *The Road Not Taken and Other Poems*. New York: Dover. (1993 reissue of *Mountain Interval*. New York: Henry Holt & Co. 1916).
Gao, S. (2013). Protective measurement: a paradigm shift in understanding quantum mechanics. Preprint.
Hardy, L. (1992). On the existence of empty waves in quantum theory. *Phys. Lett. A*, **167**, 11.

Heisenberg, W. (1930). *The Physical Principles of the Quantum Theory*. Chicago: University of Chicago Press (reprinted by Dover, New York, 1949).

Holland, P. (1994). Testing wavefunction collapse. Preprint.

Holland, P. (2006). Quantum back-reaction and the particle law of motion. *J. Phys. A: Math. Gen.*, **39**, 559.

Holland, P.R. (1993). *The Quantum Theory of Motion*. Cambridge: Cambridge University Press.

Scully, M.O. (1998). Do Bohm trajectories always provide a trustworthy physical picture of particle motion? *Phys. Scr.,* **T76**, 41.

**Figure captions**

1. A particle traversing path 1 may arrive at an otherwise inaccessible point *X* due to the action of the empty wave $\psi_2$.

2. A device for determining the path of particle *x* from the remote detection of particle *y* using quantum entanglement.

3. The path of particle *x* inferred from the excitation of detector *y* depends on the prevailing quantum state: *x* traverses path 1 in case (a) and path 2 in case (b).

4. An unsuccessful method to detect an empty wave $\psi_2$ from its effect on a system *w*.

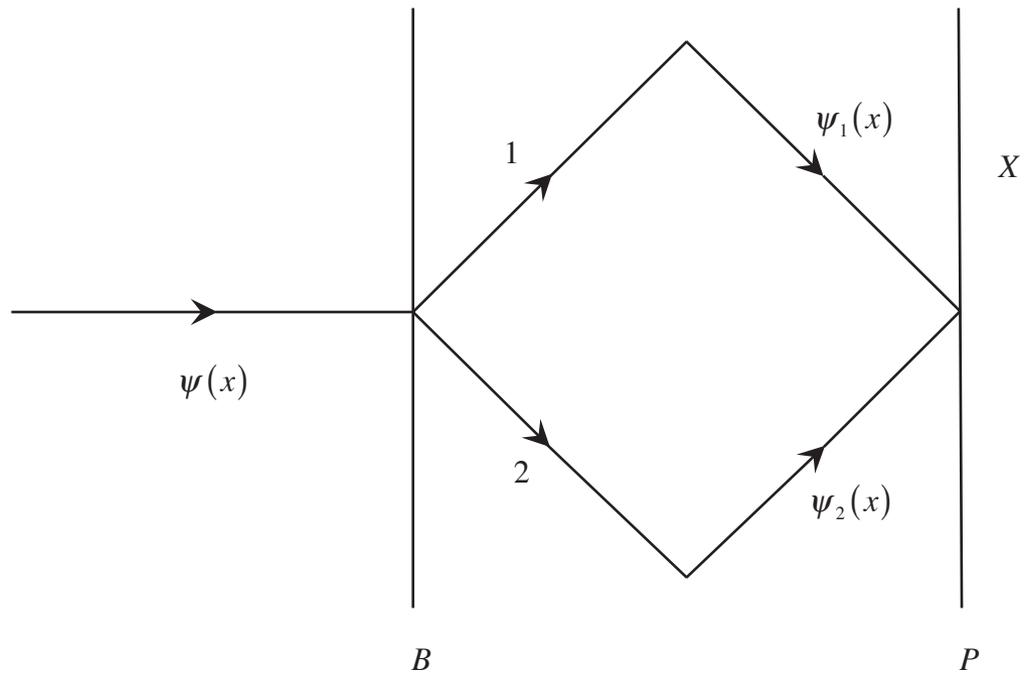

Fig. 1

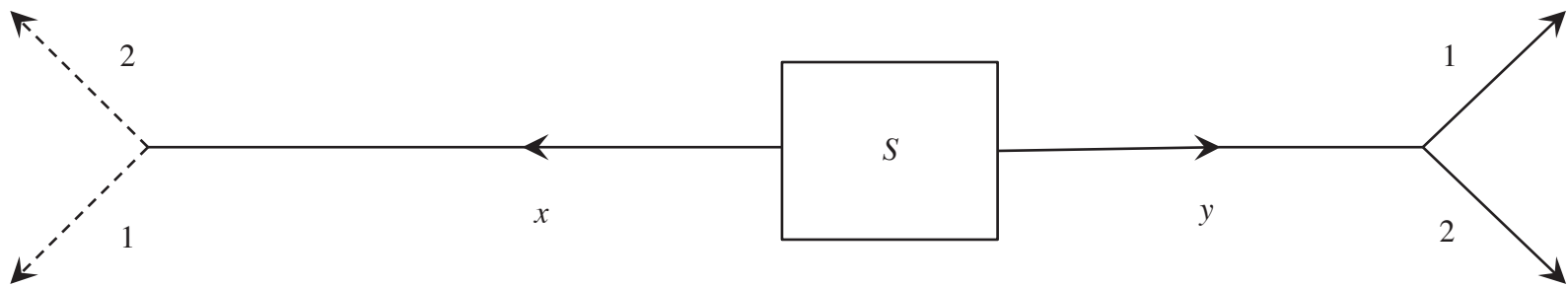

Fig. 2

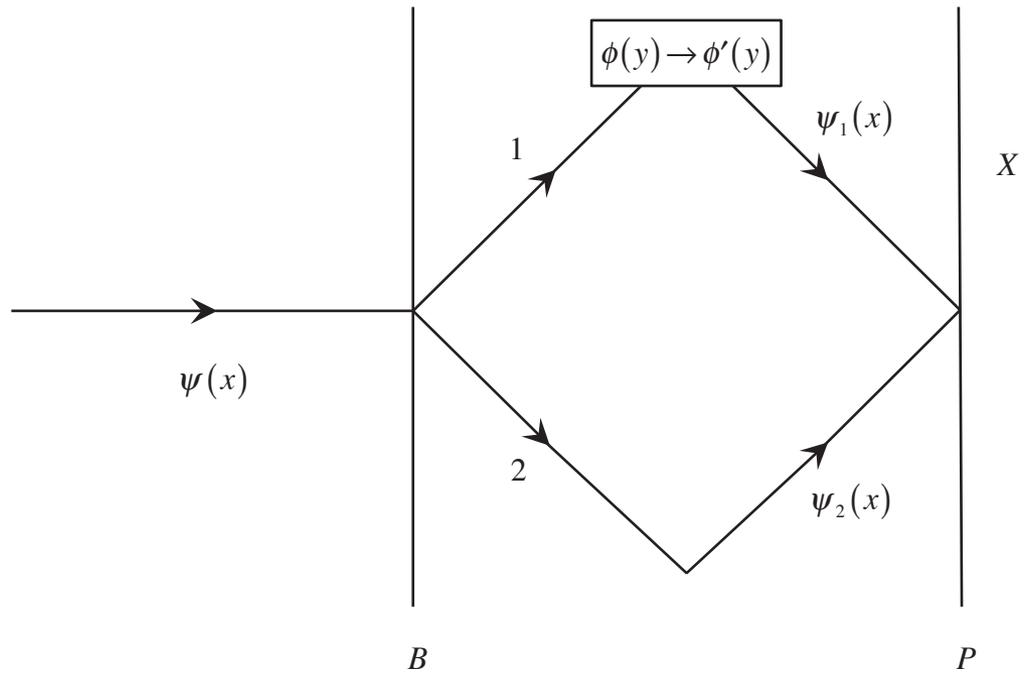

Fig. 3a

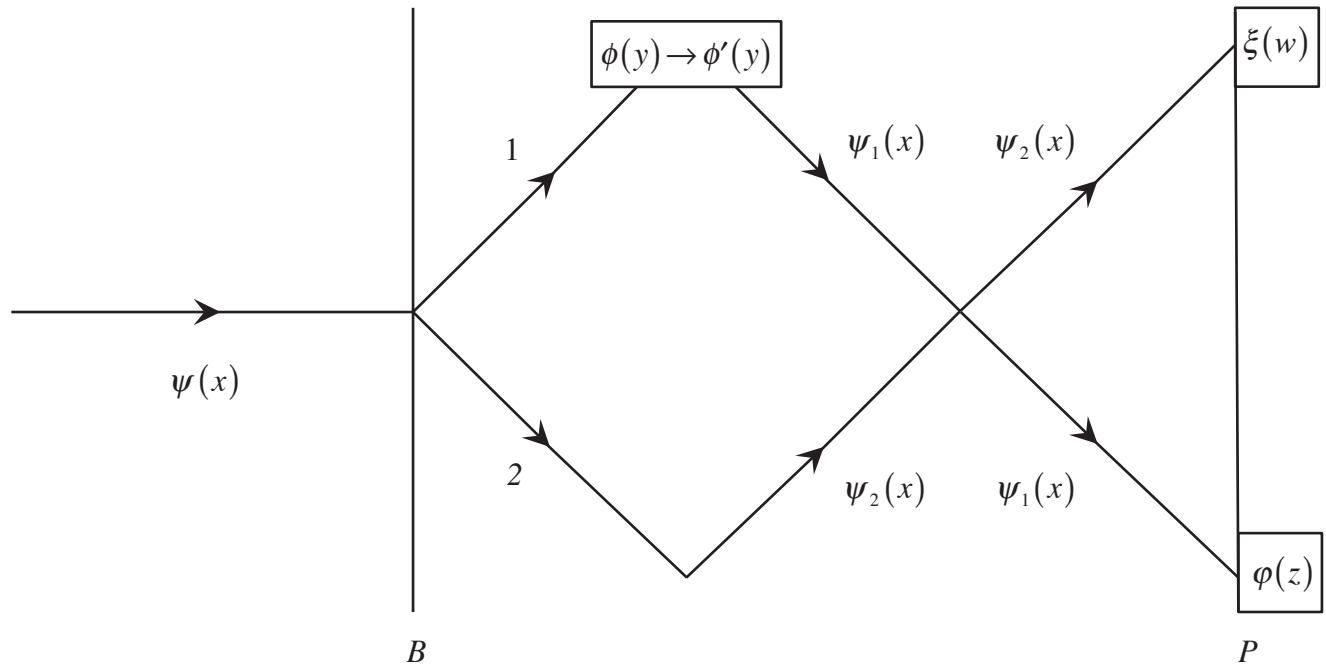

Fig. 3b

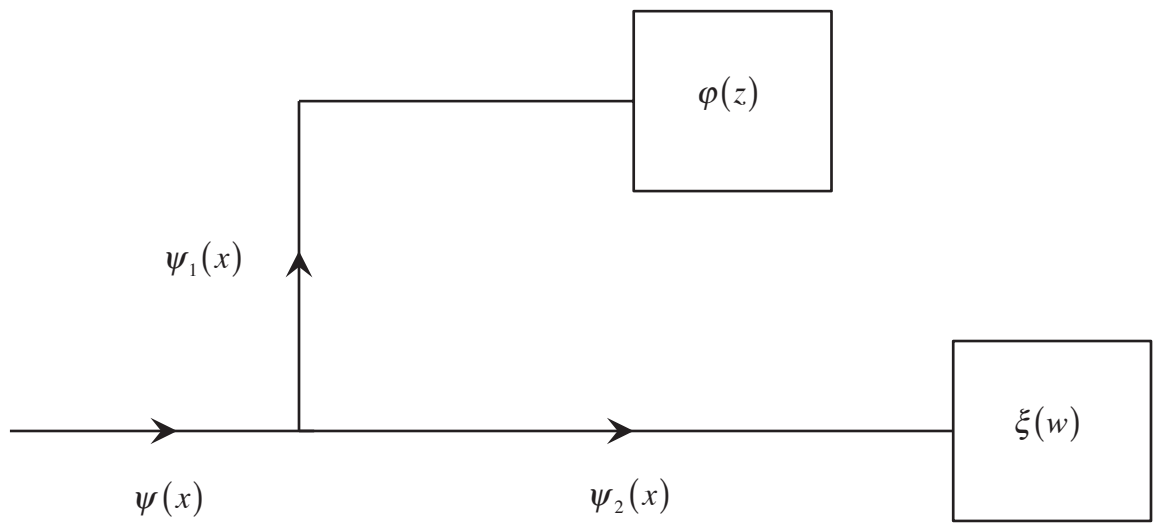

Fig. 4